\documentclass[fleqn,usenatbib]{mnras}

\usepackage{newtxtext,newtxmath}

\usepackage[T1]{fontenc}

\DeclareRobustCommand{\VAN}[3]{#2}
\let\VANthebibliography\thebibliography
\def\thebibliography{\DeclareRobustCommand{\VAN}[3]{##3}\VANthebibliography}

\usepackage{graphicx}	% Including figure files
\usepackage{amsmath}	% Advanced maths commands
\usepackage{amssymb}	% Extra maths symbols
\usepackage[utf8]{inputenc}
\usepackage[english]{babel}
\usepackage{booktabs}
\usepackage{siunitx}
\newcommand{\orcid}[1]{\href{https://orcid.org/#1}{\includegraphics[width=8pt]{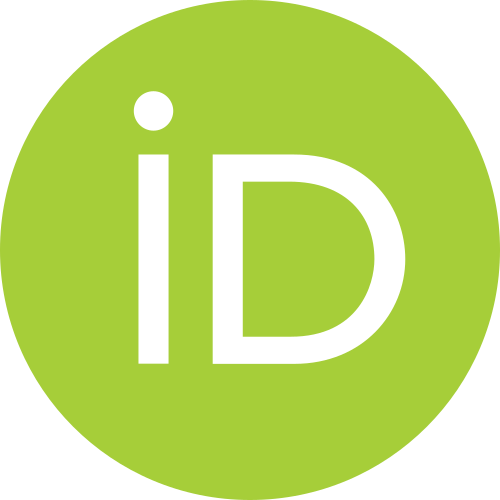}}}
\newcommand{\QR}{$Q_{\textnormal{R}}$ }
\newcommand{\QRD}{$Q_{\textnormal{RD}}^{*}$ }

\graphicspath{{./}{plots/}}

\DeclareSIUnit\Mearth{M_\oplus}
\DeclareSIUnit\Rearth{R_\oplus}
\DeclareSIUnit\erg{erg}

\title[EOS/Resolution Conspiracy]{The EOS/Resolution Conspiracy: Convergence in Proto-Planetary Collision Simulations}

\author[T. Meier et al.]{
Thomas Meier\orcid{0000-0001-9682-8563}$^{1}$\thanks{E-mail:thomas.meier5@uzh.ch},
Christian Reinhardt\orcid{0000-0002-4535-3956}$^{1}$,
Joachim Gerhard Stadel\orcid{0000-0001-7565-8622}$^{1}$
\\
$^{1}$Institute for Computational Science, University of Zurich, Winterthurerstrasse 190, 8057 Zurich, Switzerland
}

\date{Accepted XXX. Received YYY; in original form ZZZ}

\pubyear{2021}

\begin{document}
\label{firstpage}
\pagerange{\pageref{firstpage}--\pageref{lastpage}}
\maketitle

% Abstract of the paper
\begin{abstract}
We investigate how the choice of equation of state (EOS) and resolution conspire to affect the outcomes of giant impact (GI) simulations. We focus on the simple case of equal mass collisions of two Earth-like \SI{0.5}{\Mearth} proto-planets showing that the choice of EOS has a profound impact on the outcome of such collisions as well as on the numerical convergence with resolution. In simulations where the Tillotson EOS is used, impacts generate an excess amount of vapour due to the lack of a thermodynamically consistent treatment of phase transitions and mixtures. In oblique collisions this enhances the artificial angular momentum (AM) transport from the planet to the circum-planetary disc reducing the planet's rotation period over time. Even at a resolution of  $1.3 \times 10^6$ particles the result is not converged. In head-on collisions the lack of a proper treatment of the solid/liquid-vapour phase transition allows the bound material to expand to very low densities which in turn results in very slow numerical convergence of the critical specific impact energy for catastrophic disruption \QRD with increasing resolution as reported in prior work. The simulations where ANEOS is used for oblique impacts are already converged at a modest resolution of $10^5$ particles, while head-on collisions converge when they evidence the post-shock formation of a dense iron-rich ring, which promotes gravitational re-accumulation of material. Once sufficient resolution is reached to resolve the liquid-vapour phase transition of iron in the ANEOS case, and this ring is resolved, the value of \QRD has then converged.
\end{abstract}

% Select between one and six entries from the list of approved keywords.
% Don't make up new ones.
\begin{keywords}
hydrodynamics -- equation of state -- software: simulations -- planets and satellites: general -- planets and satellites: formation -- planets and satellites: terrestrial planets
\end{keywords}

%%%%%%%%%%%%%%%%%%%%%%%%%%%%%%%%%%%%%%%%%%%%%%%%%%
%%%%%%%%%%%%%%%%% BODY OF PAPER %%%%%%%%%%%%%%%%%%

\section{Introduction}
The last stage of terrestrial planet formation is characterised by extremely energetic impacts between the remaining proto-planets \citep{agnorCharacterConsequencesLarge1999}. Such giant impacts (GI) play a key role in shaping the final structure of planetary systems and therefore have been a very active branch of research over the last few decades (e.g., \citealt{benzCatastrophicDisruptionsRevisited1999,canupSimulationsLateLunarforming2004,jutziStructureAsteroidVesta2013,citronFormationPhobosDeimos2015,chauFormingMercuryGiant2018,emsenhuberSPHCalculationsMarsscale2018,kegerreisConsequencesGiantImpacts2018,dengEnhancedMixingGiant2019}). Since laboratory experiments are limited to much lower energies and have difficulties to account for large scale features like self gravity or decreasing material strength with increasing size of the target body (\citealt{jutziNumericalSimulationsImpacts2009,remingtonNumericalSimulationsLaboratoryScale2020,okamotoImpactEjectaImpact2020}), computer simulations are the dominant tool to study the outcomes of such global collisions.

One key ingredient for such simulations is an equation of state (EOS) that models the material's behaviour over the very large range of densities and temperatures involved in GI. Commonly, two different equations of state are dominantly used in the community: the Tillotson EOS \citep{tillotsonMetallicEquationsState1966} and ANEOS (ANalytic Equation Of State) \citep{thompsonImprovementsCHARTRadiationhydrodynamic1974} or its newer variant M-ANEOS \citep{meloshHydrocodeEquationState2007}. Both EOS have been used to investigate a wide range of impact conditions and problems that cover collisions between kilometre sized planetesimals (e.g., \citealt{benzCatastrophicDisruptionsRevisited1999,jutziStructureAsteroidVesta2013,gendaResolutionDependenceDisruptive2015}) over impacts on Mercury (\citealt{benzOriginMercury2007,asphaugMercuryOtherIronrich2014,chauFormingMercuryGiant2018}), Earth (\citealt{benzOriginMoonSingleimpact1987,canupSimulationsLateLunarforming2004,canupFormingMoonEarthlike2012,cukMakingMoonFastSpinning2012,lockOriginMoonTerrestrial2018,dengPrimordialEarthMantle2019}) and Mars (\citealt{marinovaMegaimpactFormationMars2008,emsenhuberSPHCalculationsMarsscale2018}) to GI in the early history of the ice giants (\citealt{kurosakiExchangeMassAngular2018,kegerreisConsequencesGiantImpacts2018,kegerreisPlanetaryGiantImpacts2019,reinhardtBifurcationHistoryUranus2020}) and the formation of the Pluto-Charon binary system \citep{canupGIANTIMPACTORIGIN2010}. Despite the popularity of both EOS, there are rather few studies that directly compare how the choice of EOS affects the outcomes of impacts. In the case of the Moon forming collision, it is found that the results are in general agreement, differences are most pronounced in the disc mass and iron content as well as the angular momentum distribution (\citealt{benzOriginMoonSingle1989,canupSimulationsLateLunarforming2004}). \citet{emsenhuberSPHCalculationsMarsscale2018} compare the two EOS in simulations that investigated GI on Mars. They confirm the general agreement found in prior work and show that for such low velocity impacts the planet's post-impact temperature distribution is very similar.

Besides the EOS, numerical parameters can also affect the outcome of GI. In the case of GI on Uranus \citet{kegerreisPlanetaryGiantImpacts2019} and \citet{reinhardtBifurcationHistoryUranus2020} find that the planet's final rotation period is resolution dependent and that convergence requires $>5 \times 10^6$ particles. \citet{hosonoUnconvergenceVerylargescaleGiant2017} study how the disc and satellite mass vary with resolution for the Moon forming GI and show that the results seem to converge at $10^6$ particles but deviate from this trend at ultra high resolutions, i.e., $>3 \times 10^7$ particles. \citet{gendaResolutionDependenceDisruptive2015} (from here on G15) investigate disruptive impacts on planetesimals in the gravity dominated regime using the Smoothed Particle Hydrodynamics (SPH) method and the Tillotson EOS and find that the critical specific impact energy for catastrophic disruption depends on the simulation's resolution. They show that in order to correctly model the transformation of impact to thermal energy in a collision (and therefore the erosion of the planetesimal) the shock wave has to be adequately resolved. While moderate resolutions of a few $10^5$ particles are within \SI{50}{\percent} of the convergence limit, convergence is predicted at $>10^8$ particles which is currently prohibitively expensive. They propose to investigate this effect under different conditions, e.g., involving larger bodies and suggest that material strength and higher impact velocities could also affect convergence. However, one key aspect, how the choice of EOS affects convergence, is not considered in G15.

We compare the two EOS, the Tillotson EOS and ANEOS, within the same numerical framework and investigate how the choice of EOS affects basic post-impact properties of the planet as well as their numerical convergence. In oblique collisions, with and without pre-impact rotation, we determine the planet's post-impact rotation period. In head-on collisions we revisit the resolution dependence of the critical specific impact energy required for catastrophic disruption found in G15 and investigate how the choice of EOS affects its value and numerical convergence. In both cases we study the collision of equal mass, differentiated proto-planets with an Earth-like composition, which is a common scenario during the last phase of terrestrial planet formation. We find that even such basic properties as the planet's rotation period and the critical specific impact energy required for disruption vary as a result of the choice the EOS. Moreover, we find that in case of ANEOS both quantities converge much faster than in the corresponding simulations where the Tillotson EOS is used which substantially reduces computational requirements for such simulations and arguably increases their level of physical realism.

The paper is structured as follows: in Section \ref{sec_methods} we discuss the two EOS and how the initial conditions are built, we then present the results and discuss our findings for both impact scenarios in Section \ref{sec_resultsanddiscussion} and give a summary and conclusions in Section \ref{sec_conclusions}.

%%%%%%%%%%%%%%%%%%%%%%%%%%%%%%%%%%%%%%%%%%%%%%%%
% Section 1
%%%%%%%%%%%%%%%%%%%%%%%%%%%%%%%%%%%%%%%%%%%%%%%%
\section{Methods}
\label{sec_methods}
All impact simulations are performed using the SPH code \textsc{Gasoline} \citep{wadsleyGasolineFlexibleParallel2004} with the modifications described in \citet{reinhardtNumericalAspectsGiant2017} and \citet{reinhardtBifurcationHistoryUranus2020}. The code is further modified with a general EOS interface that enables the use of other equations of state in GI simulations \citep{meierEquationsStateCollision2020}.\footnote{The source code of the modifications to incorporate ANEOS into \textsc{Gasoline} is available at: \citet{meierANEOSmaterial2021} and \citet{meierEOSlib2021}.}

\subsection{Equations of State}
For the present work we use two different EOS which were dominantly used in prior simulations. The Tillotson EOS was specifically developed to model hyper-velocity impacts \citep{tillotsonMetallicEquationsState1966}. The main goal was to provide a simple, analytic EOS that covers the huge range of densities and temperatures involved in GI. At low compression it matches the Mie-Gruneisen EOS (\citealt{mieZurKinetischenTheorie1903,grueneisenTheorieFestenZustandes1912}) and converges to the Thomas-Fermi limit \citep{zeldovichPhysicsShockWaves1967} for very large densities and temperatures. It also contains a simple treatment for expanded states if material is evaporated due to post-shock expansion. Despite the simple analytic form it is in good agreement with experimental data \citep{benzOriginMoonSingleimpact1986, brundageImplementationTillotsonEquation2013} and its ability to capture shocks and therefore reproduce the material's Hugoniot curve is excellent \citep{brundageImplementationTillotsonEquation2013}. However, it lacks a thermodynamically consistent treatment of mixed phases or phase transitions and the good agreement to experimental data is limited to relatively low velocity collisions where only a small fraction of the material is (partially) vaporised \citep{emsenhuberSPHCalculationsMarsscale2018}. The region, where differences to more sophisticated EOS are expected to be most pronounced are the expanded, intermediate states which the Tillotson EOS treats as a simple interpolation in pressure between a cold solid-liquid and a gas phase (e.g., \citealt{benzOriginMoonSingleimpact1986,canupSimulationsLateLunarforming2004,stewartShockPhysicsGiant2020}, see Appendix \ref{appendix_treatment_neg_pres_till} for details concerning the implementation). Furthermore, the Tillotson EOS is not thermodynamically complete and does not provide the temperature $T$. It therefore has to be estimated from
\begin{equation}
    T = \frac{u}{c_v}
\end{equation}
where $u$ is the specific internal energy and $c_v$ the specific heat capacity of the material assumed to be constant.

The ANEOS (ANalytic Equation Of State) equation of state \citep{thompsonImprovementsCHARTRadiationhydrodynamic1974} is based on fitting analytic expressions of the Helmholtz free energy in different phases of the material to experimental data. The Helmholtz free energy $F$ is expressed as the sum of three components that correspond to contributions from atomic and electronic interactions at zero temperature, a temperature-dependent part of the inter-atomic forces and electronic effects. From the Helmholtz free energy the pressure, internal energy, entropy and sound speed can be directly obtained using standard thermodynamic relations. Crucial for the present work is the thermodynamically consistent treatment of phase transitions and mixed phases, a well-known short-coming of the Tillotson EOS. In its original form ANEOS supports solid-liquid (melting), solid/liquid-vapour (vaporisation) and a single solid-solid phase transition. However, the solid-liquid and the solid-solid phase transition can not be accounted for simultaneously. Which phase transitions are implemented depends on the material. In the present work we use iron and dunite. In case of iron the solid-liquid (melting) and solid/liquid-vapour phase transitions are implemented, whereas dunite supports the solid-solid (corresponding to a change in crystallisation state from olivine to spinel at \SI{660}{\giga\pascal}, \citealt{benzOriginMoonSingle1989}) and solid/liquid-vapour transitions.

Since the EOS parameters for the same material can vary between different sources we provide them for all materials used in our simulations in Table \ref{tab_ANEOSParameters} (iron and dunite for ANEOS) and Table \ref{tab_TillotsonParameters} (iron and granite for the Tillotson EOS). In case of the Tillotson EOS it is possible to calculate the pressure and sound speed from the analytic expressions (see \citealt{reinhardtNumericalAspectsGiant2017} for details) at each step of the simulations. For ANEOS this is not possible because EOS function calls are slow and the \texttt{Fortran} code cannot be evaluated in parallel without major modifications. Therefore, we generate EOS tables with $1401\times1201$ grid points with logarithmic spacing in density and temperature for all the materials involved in a simulation. On this grid linear interpolation is performed for each EOS call. Due to the fine grid, the interpolated values agree very well ($<\SI{1}{\percent}$) with the values calculated directly from ANEOS. Larger deviations are only found in a small region where the ANEOS code itself interpolates \citep{thompsonImprovementsCHARTRadiationhydrodynamic1974} due to different choices of the interpolation grid points.
\label{sec:interpolation}

\begin{table}
    \centering
    \begin{tabular}{ccc}
        \toprule
        &Iron&Dunite\\
        \midrule
        V1&1&3\\
        V2&4&4\\
        V3&\SI{7.85}{\gram\per\centi\meter\tothe3}&\SI{3.32}{\gram\per\centi\meter\tothe3}\\
        V6&\SI{1.45e12}{\gram\per\centi\meter\per\second\tothe2}&\SI{-6.6e5}{\centi\meter\per\second}\\
        V7&1.690&0.82\\
        V8&\SI{-0.0400}{\electronvolt}&\SI{0.057}{\electronvolt}\\
        V9&0&9.86\\
        V10&2&2\\
        V11&\SI{8.200e10}{\erg\per\gram}&\SI{2.e11}{\erg\per\gram}\\
        V12&\SI{0.15588}{\electronvolt}&\SI{0.19}{\electronvolt}\\
        V18&0&\SI{4.65}{\gram\per\centi\meter\tothe3}\\
        V19&0&\SI{4.9}{\gram\per\centi\meter\tothe3}\\
        V20&0&\SI{6.6e11}{\erg\per\centi\meter\tothe3}\\
        V21&0&\SI{3.5e12}{\erg\per\centi\meter\tothe3}\\
        V22&0&\SI{1.3e13}{\erg\per\centi\meter\tothe3}\\
        V23&\SI{2.471e9}{\erg\per\gram}&0\\
        V24&0.955&0\\
        Z(1), f(1)&26, 1&8, 0.571\\
        Z(2), f(2)&&12, 0.286\\
        Z(3), f(3)&&14, 0.143\\
        \toprule
    \end{tabular}
    \caption{ANEOS parameters used in the simulations. In the present work we use iron from \citet{emsenhuberSPHCalculationsMarsscale2018} and dunite from \citet{benzOriginMoonSingle1989}. For an explanation of the different parameters, see \citet{thompsonANEOSAnalyticEquations1990}.}
    \label{tab_ANEOSParameters}
\end{table}
\begin{table}
    \centering
    \begin{tabular}{ccc}
        \toprule
        &Iron&Granite\\
        \midrule
        $a$&\SI{0.5}{}&\SI{0.5}{}\\
        $b$&\SI{1.5}{}&\SI{1.3}{}\\
        $u_0$&\SI{9.5e10}{\erg\per\gram}&\SI{1.6e11}{\erg\per\gram}\\
        $\rho_0$&\SI{7.86}{\gram\per\centi\meter\tothe3}&\SI{2.7}{\gram\per\centi\meter\tothe3}\\
        $A$&\SI{1.279e12}{\erg\per\centi\meter\tothe3}&\SI{1.8e11}{\erg\per\centi\meter\tothe3}\\
        $B$&\SI{1.05e12}{\erg\per\centi\meter\tothe3}&\SI{1.8e11}{\erg\per\centi\meter\tothe3}\\
        $u_s$&\SI{1.42e10}{\erg\per\gram}&\SI{3.5e10}{\erg\per\gram}\\
        $u_{s2}$&\SI{8.45e10}{\erg\per\gram}&\SI{1.8e11}{\erg\per\gram}\\
        $\alpha$&\SI{5.0}{}&\SI{5.0}{}\\
        $\beta$&\SI{5.0}{}&\SI{5.0}{}\\
        $c_V$&\SI{0.449e7}{\erg\per\gram\per\kelvin}&\SI{0.79e7}{\erg\per\gram\per\kelvin}\\
        \toprule
    \end{tabular}
    \caption{Tillotson EOS parameters used in the simulations. In the present work, we use the parameters for granite from \citet{benzOriginMoonSingleimpact1986} and for iron from \citet{benzOriginMoonSingleimpact1987}.}
    \label{tab_TillotsonParameters}
\end{table}

\subsection{Initial conditions}
\label{sec_IC}
The pre-impact models of the proto-planets are created using \textsc{ballic} \citep{reinhardtNumericalAspectsGiant2017} with improvements for multi-component models described in \citet{chauFormingMercuryGiant2018} and \citet{reinhardtBifurcationHistoryUranus2020}. The proto-planets involved in the collisions are identical and have a mass of \SI{0.525}{\Mearth} and an Earth-like composition with an iron core (\SI{33}{\percent}) and a rocky mantle (\SI{67}{\percent}). The surface temperature is chosen to be $T_s=\SI{1000}{\kelvin}$. Since there is no identical material match between Tillotson and ANEOS for the mantle, we use dunite \citep{benzOriginMoonSingle1989} for ANEOS and granite \citep{benzOriginMoonSingleimpact1986} in case of the Tillotson EOS. While the EOS parameters differ slightly, e.g., the materials density at the reference state $\rho_0$, the two materials show very similar properties. Most importantly the shock Hugoniot in the condensed states and therefore the peak shock pressure and temperature are in a good agreement for moderate compression. At higher compression the solid-solid phase transition in dunite, which is not accounted for in granite, leads to larger differences. The proto-planets are sampled with \SI{5e3}{} to \SI{1e6}{} particles, or a total number of particles varying between \SI{1e4}{} to \SI{2e6}{} for each collision simulation.

For the impacts where the proto-planets have an initial rotation prior to the collision, the models are assigned a uniform rotation as described in \citet{timpeMachineLearningApplied2020}.
The angular velocity for all rotating models is set as
\begin{equation}
  \omega=0.9\omega_{\textnormal{break}}=0.9\sqrt{\pi G\bar{\rho}h_{\textnormal{crit}}}
\end{equation}
where $\bar{\rho}$ is the mean density of the body and $h_{\textnormal{crit}}=0.44931$ is given by MacLaurin's formula for the maximum angular velocity (\citealt{chandrasekharEllipsoidalFiguresEquilibrium1969,ansorgUniformlyRotatingAxisymmetric2003}). The proto-planets are therefore rotating close to the critical rotation period for breakup, thereby maximising any possible additional effects due to rotation. The radii, gravitational binding energies, escape velocities and mean densities of the resulting bodies as well as their rotation periods, angular velocities and spin angular momentum are shown in Table \ref{tab_BodyProperties}.

\begin{table}
    \centering
    \begin{tabular}{ccc}
        \toprule
        &Tillotson&ANEOS\\
        \midrule
        $R_{c}$&\SI{0.438}{\Rearth}&\SI{0.440}{\Rearth}\\
        $R_{p}$&\SI{0.830}{\Rearth}&\SI{0.833}{\Rearth}\\
        $E$&\SI{8.401e31}{\joule}&\SI{8.278e31}{\joule}\\
        $v_{\textnormal{esc}}$&\SI{8.883}{\kilo\meter\per\second}&\SI{8.866}{\kilo\meter\per\second}\\
        $\bar{\rho}$&\SI{5.041}{\gram\per\centi\meter\tothe3}&\SI{4.984}{\gram\per\centi\meter\tothe3}\\
        \midrule
        $R_{\textnormal{rot}}$&\SI{0.989}{\Rearth}&\SI{0.961}{\Rearth}\\
        $P$&\SI{2.81}{\hour}&\SI{2.83}{\hour}\\
        $\omega$&\SI{6.202e-4}{\per\second}&\SI{6.166e-4}{\per\second}\\
        $L$&\SI{2.399e34}{\kilo\gram\meter\tothe2\per\second}&\SI{2.370e34}{\kilo\gram\meter\tothe2\per\second}\\
        \toprule
    \end{tabular}
    \caption{The pre-impact properties of the colliding proto-planets: Radius of the core-mantle boundary $R_{c}$, total radius $R_{p}$, gravitational binding energy $E$, escape velocity $v_{\mbox{esc}}$ and mean density $\bar{\rho}$ of the spherical, non-rotating bodies and semi-major axis $R_{\textnormal{rot}}$, rotation period $P$, angular velocity $\omega$ and angular momentum $L$ if the proto-planets are assigned a pre-impact rotation. Despite the different reference densities of the mantle materials the radii and therefore the mutual escape velocity and collision geometries are very similar.}
    \label{tab_BodyProperties}
\end{table}

\subsubsection{Oblique collisions}
In the first impact scenario we investigate oblique collisions between the two proto-planets with and without pre-impact rotation at different resolutions. The goal is to determine the final planet's rotation period. The trajectory of the colliding bodies is defined by the impact parameter $b_{\textnormal{inf}}=0.95$ and velocity $v_{\textnormal{inf}}=0.5$ at infinity. The procedure to calculate the initial positions and velocities from these values is described in detail in appendix \ref{appendix_oblique_initial_condition}. With this trajectory, that has an angular momentum (AM) of $L=\SI{1.21e35}{\kilogram\meter\tothe2\per\second}$, four different configurations of the spinning and non-spinning bodies are simulated:

\begin{itemize}
    \item NR: Two non-rotating bodies are used, resulting in a total angular momentum of $L_{\textnormal{NR}}=\SI{1.21e35}{\kilogram\meter\tothe2\per\second}$
    \item UU: Two rotating bodies are used such that both spin vectors are parallel to the orbital angular momentum, resulting in a total angular momentum of $L_{\textnormal{UU}}=\SI{1.69e35}{\kilogram\meter\tothe2\per\second}$
    \item DD: The two rotating bodies are placed such that both spin vectors are anti-parallel to the orbital angular momentum, resulting in a total angular momentum of $L_{\textnormal{DD}}=\SI{0.73e35}{\kilogram\meter\tothe2\per\second}$
    \item DU: The spin vector of the projectile is parallel and the spin vector of the target is anti-parallel to the orbital angular momentum, cancelling each other, and resulting in the same total angular momentum as in the NR case $L_{\textnormal{DU}}=\SI{1.21e35}{\kilogram\meter\tothe2\per\second}$
\end{itemize}

\subsubsection{Head-on collisions}
The second impact scenario consists of head-on ($b_{\inf} = 0$) collisions between the two (non-rotating) proto-planets. We determine the critical specific impact energy for catastrophic disruption, \QRD, at which the remaining bound material comprises \SI{50}{\percent} of the total colliding mass. The specific impact energy in the centre of mass frame is defined as
\begin{equation}
    Q_R=\frac{1}{2} \frac{\mu v_{\textnormal{imp}}^2}{M_{\textnormal{tot}}}
\end{equation}
where $v_{\textnormal{imp}}$ is the impact velocity, $\mu$ is the reduced mass and $M_{\textnormal{tot}}$ is the total colliding mass. In the case where both colliding proto-planets have the same mass this reduces to
\begin{equation}
    Q_R=\frac{1}{8} v_{\textnormal{imp}}^2
\end{equation}
from which $v_{\textnormal{imp}}$ is calculated for each collision. For the range of \QR investigated in this study, this results in impact velocities ranging from 1 to 3 times the mutual escape velocity of the system
\begin{equation}
    v_{\textnormal{esc,sys}}= \sqrt{\frac{2G(M_{\textnormal{targ}}+M_{\textnormal{proj}})}{R_{\textnormal{targ}}+R_{\textnormal{proj}}}}
\end{equation}
where $M_\textnormal{targ}$, $M_\textnormal{proj}$, $R_\textnormal{targ}$ and $R_\textnormal{proj}$ are the mass and radius of the target and the projectile respectively and $G$ is the gravitational constant. This further simplifies to
\begin{equation}
    v_{\textnormal{esc,sys}} = \sqrt{\frac{2 G M}{R}}
\end{equation}
when the colliding proto-planets have equal masses $M$ and radii $R$.

\subsection{Analysis}
All impact simulations result in a gravitationally bound remnant that is distinguished from the ejecta using the group finder \textsc{skid}\footnote{The source code is available at: \url{https://github.com/N-BodyShop/skid}.} \citep{stadelNBodyShopSkid2017}. The oblique collisions deposit material in orbit that forms a circum-planetary disc. In this case the gravitationally bound material is therefore further divided into a central dense region that we refer to as planet and a disc of orbiting material. This is achieved using the iterative procedure described in \citet{reinhardtBifurcationHistoryUranus2020}. All bound material which has an orbit intersecting the planet "surface" defined by its mass and the assumed mean density of $\SI{5.6}{\gram\per\centi\meter\tothe3}$ belongs to the planet, the rest belongs to the disc. The rotation period of the planet is then calculated from the median of the angular velocities of all particles classified as belonging to the planet.

For the head-on collisions the critical specific impact energy \QRD required to disrupt and gravitationally disperse half of the total colliding mass is calculated following G15. First, the mass of the gravitationally bound group of particles is obtained from \textsc{skid} for different specific impact energies \QR for a given EOS and resolution. Then \QRD is calculated as in G15 by linear interpolation between the two data points that bracket the specific impact energy where exactly half the total colliding mass remains bound.

%%%%%%%%%%%%%%%%%%%%%%%%%%%%%%%%%%%%%%%%%%%%%%%%
% Section 2
%%%%%%%%%%%%%%%%%%%%%%%%%%%%%%%%%%%%%%%%%%%%%%%%
\section{Results and Discussion}
\label{sec_resultsanddiscussion}
We focus on two simple but generally important collision outcomes in order to compare the two EOS. We perform 32 simulations to investigate the rotation period in an oblique impact and 96 simulations to determine \QRD for each EOS. This means that the suite of simulations presented in this paper consists of over 250 impact simulations.

\subsection{Oblique collision}
\begin{figure}
    \includegraphics[width=\columnwidth]{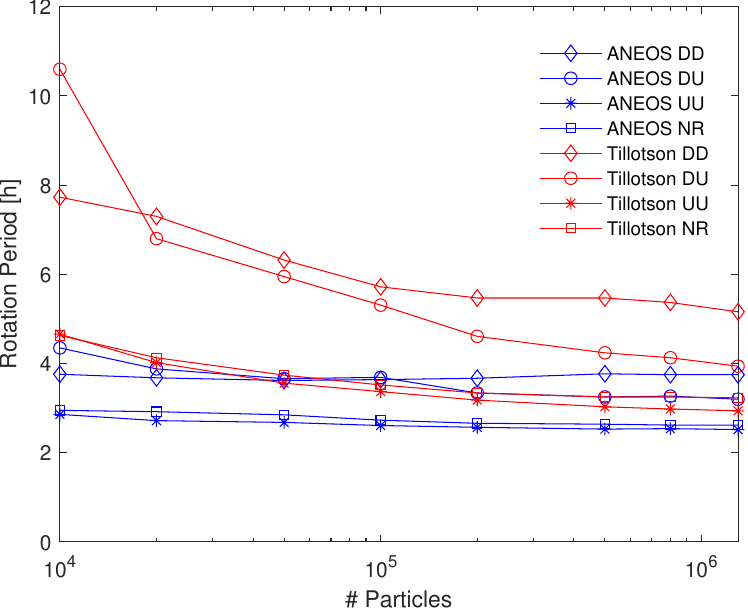}
    \caption{The resolution dependence of the planet's inferred rotation period for the different oblique impacts \SI{169}{\hour} after the collision. The colours represent the two EOS (Tillotson EOS: red, ANEOS: blue). The symbols represent the different angular momentum configurations (DU: circle, UU: star, DD: diamond and NR: square). As in prior work the results obtained with the Tillotson EOS show a strong resolution dependence while in case of ANEOS the rotation period, with exception of the DU case, has essential already converged for $10^4$ particles.}
    \label{fig_xResolution_NB_rotPeriod}
\end{figure}
\begin{figure}
    \centering
    \includegraphics[width=0.95\columnwidth]{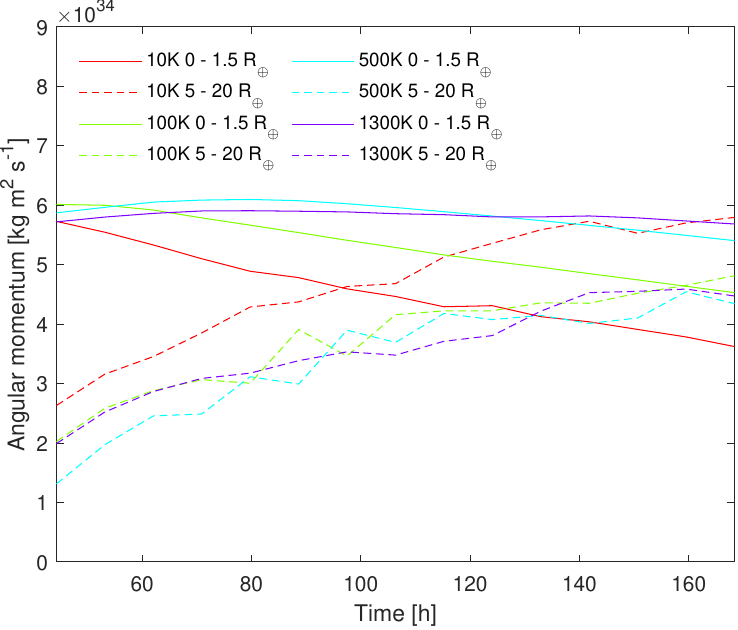}\\
    \includegraphics[width=0.95\columnwidth]{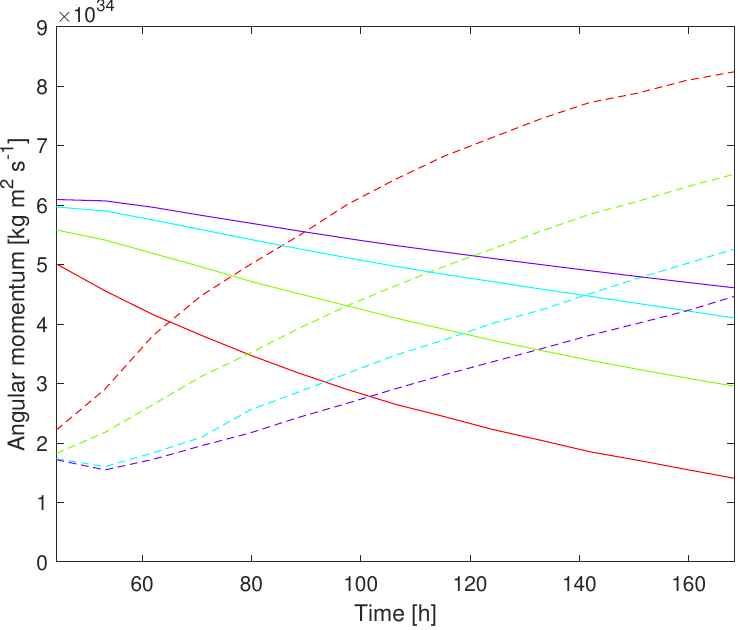}
    \caption{The time evolution of the total angular momentum in two cylindrical bins for the DU simulations for ANEOS (top) and Tillotson (bottom). The first bin (solid lines) ranges from \SIrange{0}{1.5}{\Rearth} and contains the planet and the innermost part of the disk. The second bin (dashed lines) covers most of the disk and ranges from \SIrange{5}{20}{\Rearth}. The colours correspond to different resolutions (see legend). The plot starts at $t$ = \SI{44.25}{\hour} after the impact, after the two colliding bodies have converged into a central planet and a circum-planetary disk. The initial angular momentum deposited in the planet for a given EOS is similar for all resolutions. In case of the Tillotson EOS this initial difference is much more pronounced and the angular momentum increases with resolution. The ANEOS simulations are much closer even at the lowest resolution and show no such trend. During the later evolution in all cases the total angular momentum is conserved but some of it is transported from the planet to the disk due to artificial viscosity inherent in SPH, the amount decreasing with increasing resolution. However, in case of the Tillotson EOS the relative loss of angular momentum is much larger than for ANEOS and the results have not converged even at $1.3 \times 10^6$ particles.
    }
    \label{fig_AngMomTransfer_all}
\end{figure}
\begin{figure*}
\centering
\begin{tabular}{cc}
     \includegraphics[width=0.45\linewidth]{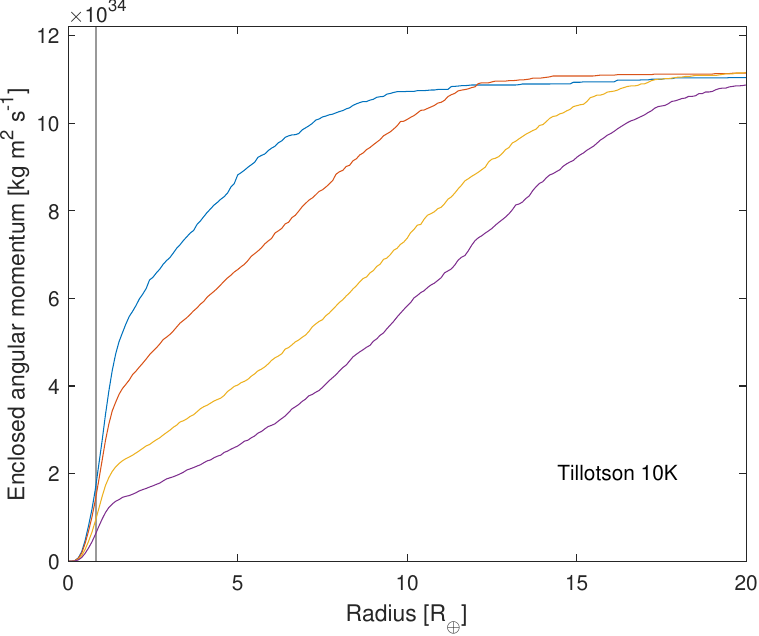} &
     \includegraphics[width=0.45\linewidth]{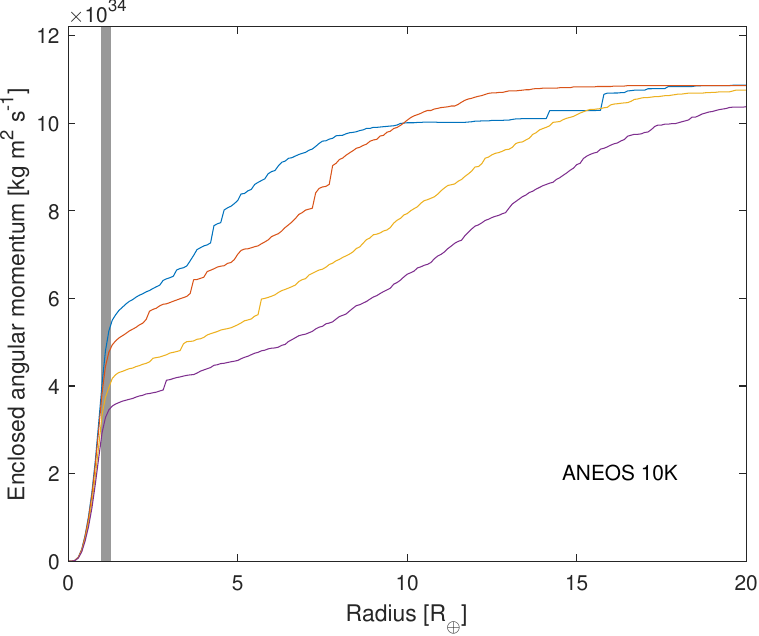}  \\
     \includegraphics[width=0.45\linewidth]{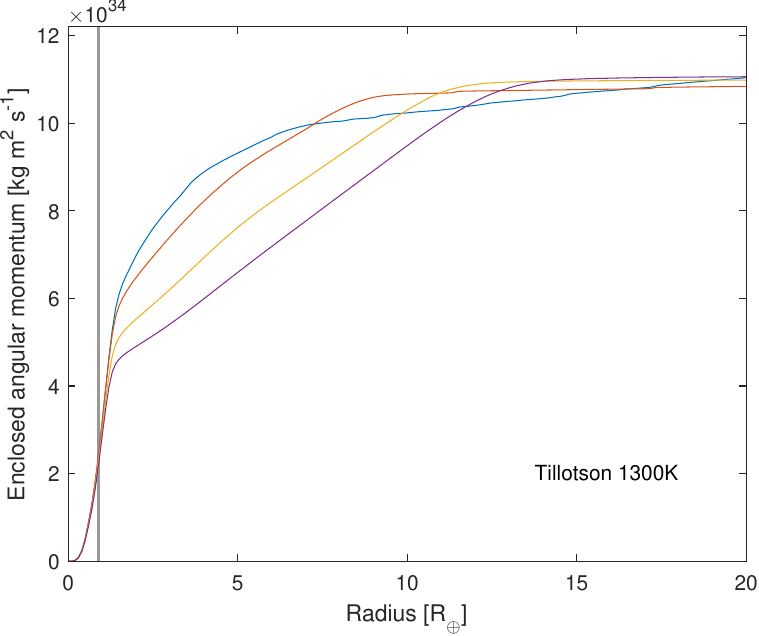} &
     \includegraphics[width=0.45\linewidth]{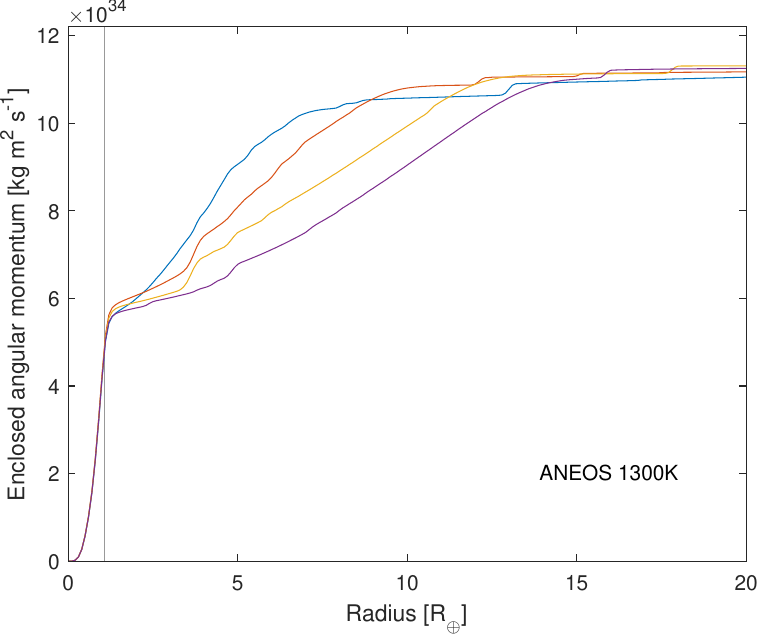}
\end{tabular}
    \caption{The enclosed angular momentum profiles for the DU case of the oblique impacts at different simulation times. Shown are the results for the Tillotson EOS (left) and ANEOS (right) at different resolutions (top: $10^4$ and bottom: $1.3 \times 10^6$ particles) \SI{1.85}{\day} (blue), \SI{2.95}{\day} (orange), \SI{5.17}{\day} (yellow) and \SI{7.01}{\day} (purple) after the impact. The shaded grey area shows the maximum radius at which particles are in the condensed state, i.e., have densities larger than the reference density. For the ANEOS simulations, most of the planet (showing a parabolic enclosed angular momentum curve consistent with solid body rotation) is in the condensed state, while in the Tillotson simulations a substantial amount of the planet is in the expanded state and deviates from solid body rotation. In the Tillotson cases a large amount of angular momentum is transferred from the planet and the inner part of the disk to the outer part of the disk over time. This effect decreases with increasing resolution but is always much larger than in the ANEOS cases. While in the ANEOS cases angular momentum is transported from the inner to the outer disk too, the planet's angular momentum is almost constant at all resolutions.}
    \label{fig_enclosed_angular_momentum}
\end{figure*}
\begin{figure*}
\centering
\begin{tabular}{cc}
     \includegraphics[width=0.45\linewidth]{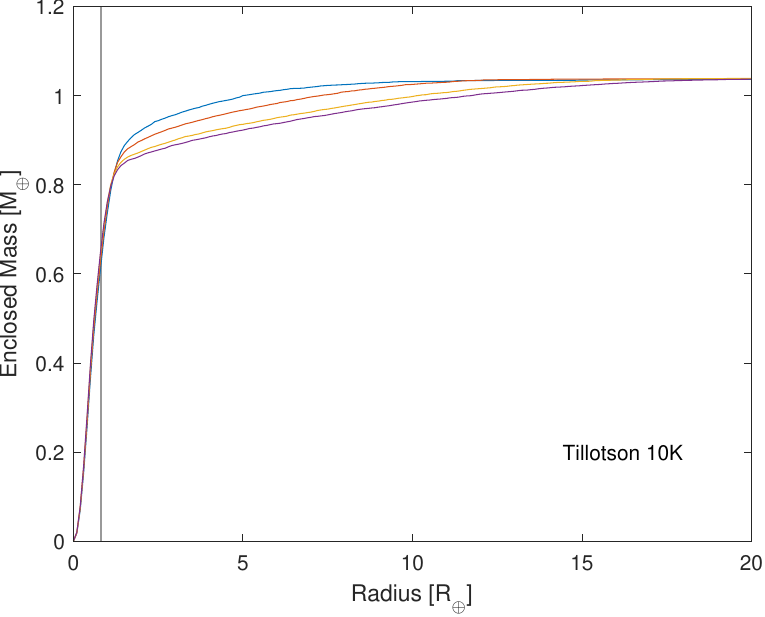} &
     \includegraphics[width=0.45\linewidth]{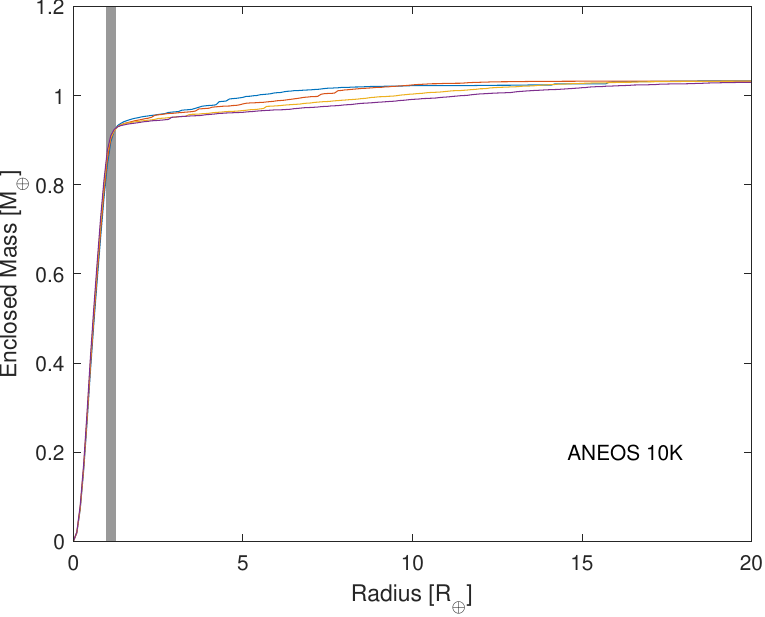}  \\
     \includegraphics[width=0.45\linewidth]{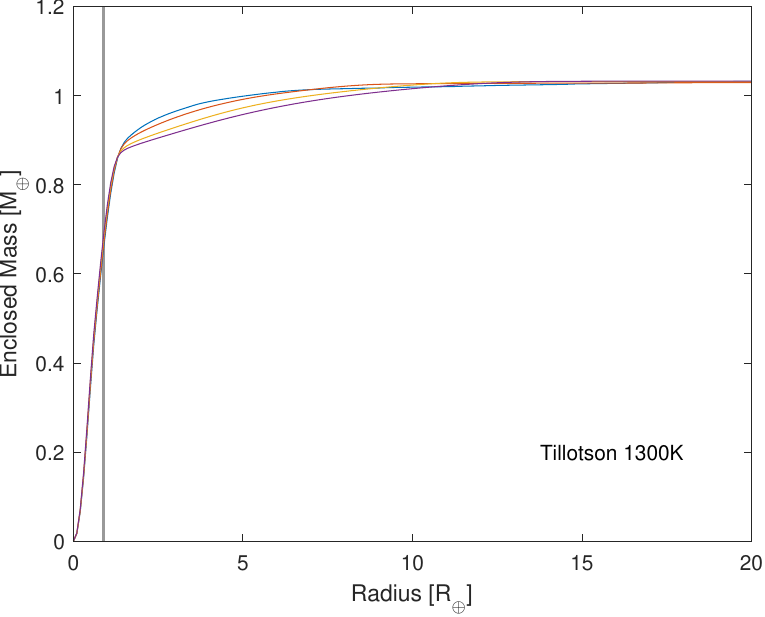} &
     \includegraphics[width=0.45\linewidth]{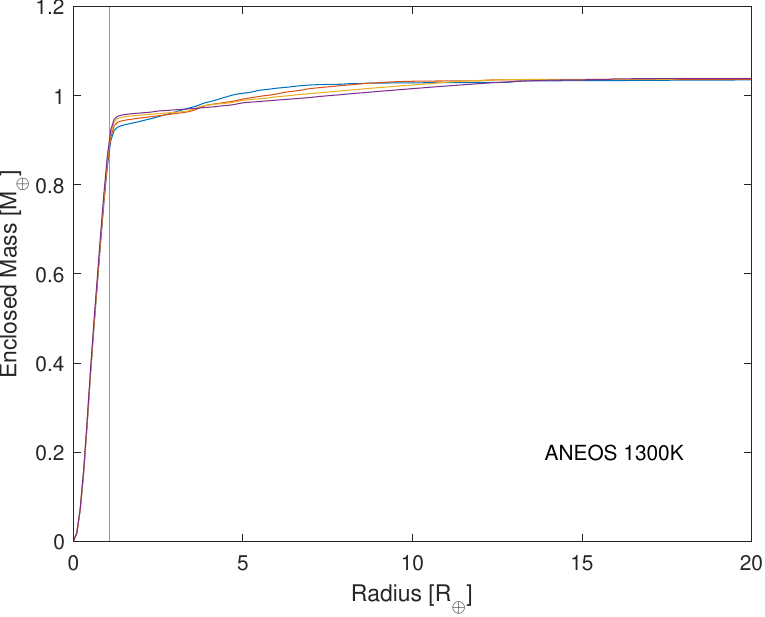}
\end{tabular}
    \caption{The enclosed mass profiles corresponding to the AM profiles for the DU case of the oblique impacts shown in Figure \ref{fig_enclosed_angular_momentum}. Due to the angular momentum transport described in Figure \ref{fig_enclosed_angular_momentum} mass is transferred from the planet and inner disk to the outer regions of the disk in all simulations. Increasing the resolution decreases the angular momentum transport and therefore the outflow of mass. For the ANEOS simulations, most of the planet is in a condensed state, i.e., the material has a density larger the reference density. In case of the Tillotson EOS a considerable amount of the material that belongs to the planet is in an expanded, low density state. One can also see that in the Tillotson simulations much more mass is transferred outwards over time.
    }
    \label{fig_enclosed_mass}
\end{figure*}

All oblique collisions result in a fast rotating planet, a circum-planetary disc and some material being ejected. Shortly after the collision, the rotation periods of the planets for both ANEOS and the Tillotson EOS show a general agreement for all resolutions but start to deviate during the later evolution. In Figure \ref{fig_xResolution_NB_rotPeriod}, the rotation period of the post-impact planet \SI{169}{\hour} after the impact is shown as a function of the number of particles in the simulation for both EOS. The rotation periods inferred from the ANEOS simulations are generally lower than the ones of the corresponding simulations where the Tillotson EOS was used. For given impact conditions the orbital angular momentum is identical for both EOS and the spin angular momentum agree within \SI{1.5}{\percent}. Since the contribution of the spin angular momentum to the initial angular momentum of the collision is small, these observed differences in rotation period are caused by the EOS. Furthermore, the rotation period varies with resolution in the Tillotson EOS runs which is consistent with prior findings (\citealt{kegerreisPlanetaryGiantImpacts2019,reinhardtBifurcationHistoryUranus2020}). However, for the ANEOS simulations we observe that the final rotation period shows little resolution dependence and, with exception of the DU case, has essentially already converged at $10^4$ particles. 

In Figure \ref{fig_AngMomTransfer_all} the time evolution of the total angular momentum for the DU case is shown. We choose the DU case because the difference in rotation period between the two EOS is most pronounced, especially at lower resolutions. The bound material is divided into two cylindrical bins. The first bin extends from the centre of the planet to \SI{1.5}{\Rearth} and contains the planet and the inner most part of the circum-planetary disc. The second bin ranges from \SI{5}{\Rearth} to \SI{20}{\Rearth} and contains the outer disc. In case of the Tillotson EOS the initial amount of AM deposited in the planet increases with resolution and the difference between the lowest and highest resolution is $\sim\SI{16}{\percent}$. This difference decreases with increasing resolution. For ANEOS on the other hand the planet's initial AM shows little variation with resolution. For all simulations, we observe that over time AM is transported from the planet and inner part of the disc to the outer region of the disc, which reduces with increasing resolution. This effect is most pronounced in the case of the Tillotson simulations where the planet loses about \SI{70}{\percent} of its initial AM at the lowest resolution. Correspondingly, the rotation period in this case is the largest in Figure \ref{fig_xResolution_NB_rotPeriod}. For ANEOS less AM is transported to the outer disc and the final rotation period of the planet is barely affected by resolution for all simulations.

We furthermore observe that the physical state of the material in the planet differs greatly depending on which EOS is used. In case of ANEOS, most of the material belonging to the planet is in the condensed states and has densities larger than the reference density $\rho_0$ while in the case of the Tillotson EOS, a large fraction of the particles that belong to the planet are in the intermediate expanded states. This in turn results in large deviations from the median (solid body) angular velocity in the outer regions of the planet. Equally, the larger pressure gradients in the expanded state produce stronger deviations from Keplerian rotation and we do not observe a clear transition from solid body to Keplerian rotation in case of the Tillotson EOS.

The initial spin of the proto-planets also affects the amount of vapour generated in the collision. The final planet and disc vapour fraction increases in the following order: NR (lowest), UU, DU, DD (highest). Correspondingly, the differences in rotation period between ANEOS and the Tillotson EOS shown in Figure \ref{fig_xResolution_NB_rotPeriod} increase in this order and are most pronounced for the DU and DD cases. This indicates a clear connection between the amount of vapour contained in the planet and the AM transport to the outer disc.

Figures \ref{fig_enclosed_angular_momentum} and \ref{fig_enclosed_mass} show the evolution of the enclosed mass and angular momentum over time of the DU case for different resolutions. As mentioned above, in the case of ANEOS the planet is mostly in the condensed state and we observe a sharp transition from the planet to the disc. The simulations that were performed using the Tillotson EOS exhibit a very smooth transition both in mass and AM from the planet to the disc because a large fraction (up to \SI{60}{\percent}) of the planet is in the expanded state. We also observe that for the Tillotson EOS the circum-planetary disc is up to \SI{30}{\percent} more massive compared to the ANEOS simulations which is consistent with prior findings (\citealt{benzOriginMoonSingle1989,canupSimulationsLateLunarforming2004}). As in Figure \ref{fig_AngMomTransfer_all} we find that AM, and corresponding mass, is transported from the planet to the disc for both EOS. However, there is no mass and very little AM transported outwards from the region of the planet where the material is in the condensed state. In the case of ANEOS, where most of the planet is condensed, the AM loss from the first to the second bin shown in Figure \ref{fig_AngMomTransfer_all} is therefore likely due to a flow of mass and AM from the inner disc rather than the planet. This explains, why the rotation period shows little variation with resolution and pre-impact spin. The large amount of AM lost from the planet to the disc in the Tillotson EOS simulations therefore seems to be related to the excess amount of vapour produced due to the lack of a more sophisticated treatment of the liquid-vapour and vapour phase.

\subsection{Head-on collisions}
\label{sec_resultsanddiscussion_headon}
\begin{figure}
    \includegraphics[width=\columnwidth]{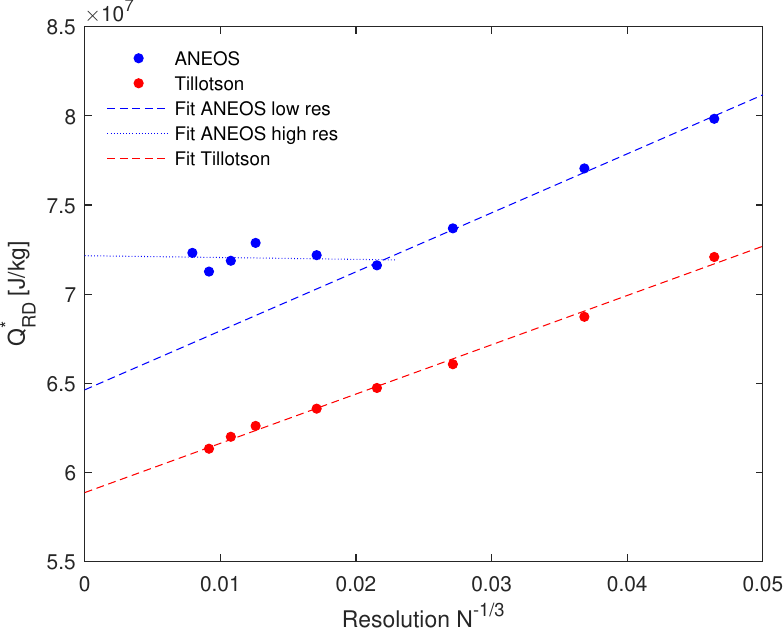}
    \caption{
    The dependence of \QRD on the number of SPH particles in the simulation with respect to the inverse of the number of particles per side $N^{-1/3}$. The values for the Tillotson EOS (red circles) show a linear dependence (linear fit shown as dashed red line) on $N^{-1/3}$ for all resolutions as in G15. For the ANEOS simulations (blue circles), this is only true for lower resolutions. For the higher resolutions ($N>10^5$), the inferred value of \QRD seems to have converged and scatters around the final value (linear fit as dotted blue line). For the values of the fitting parameters see the second paragraph of Section \ref{sec_resultsanddiscussion_headon}.
    }
    \label{fig_QDstarN-1_3}
\end{figure}
\begin{figure*}
     \includegraphics[width=0.75\linewidth]{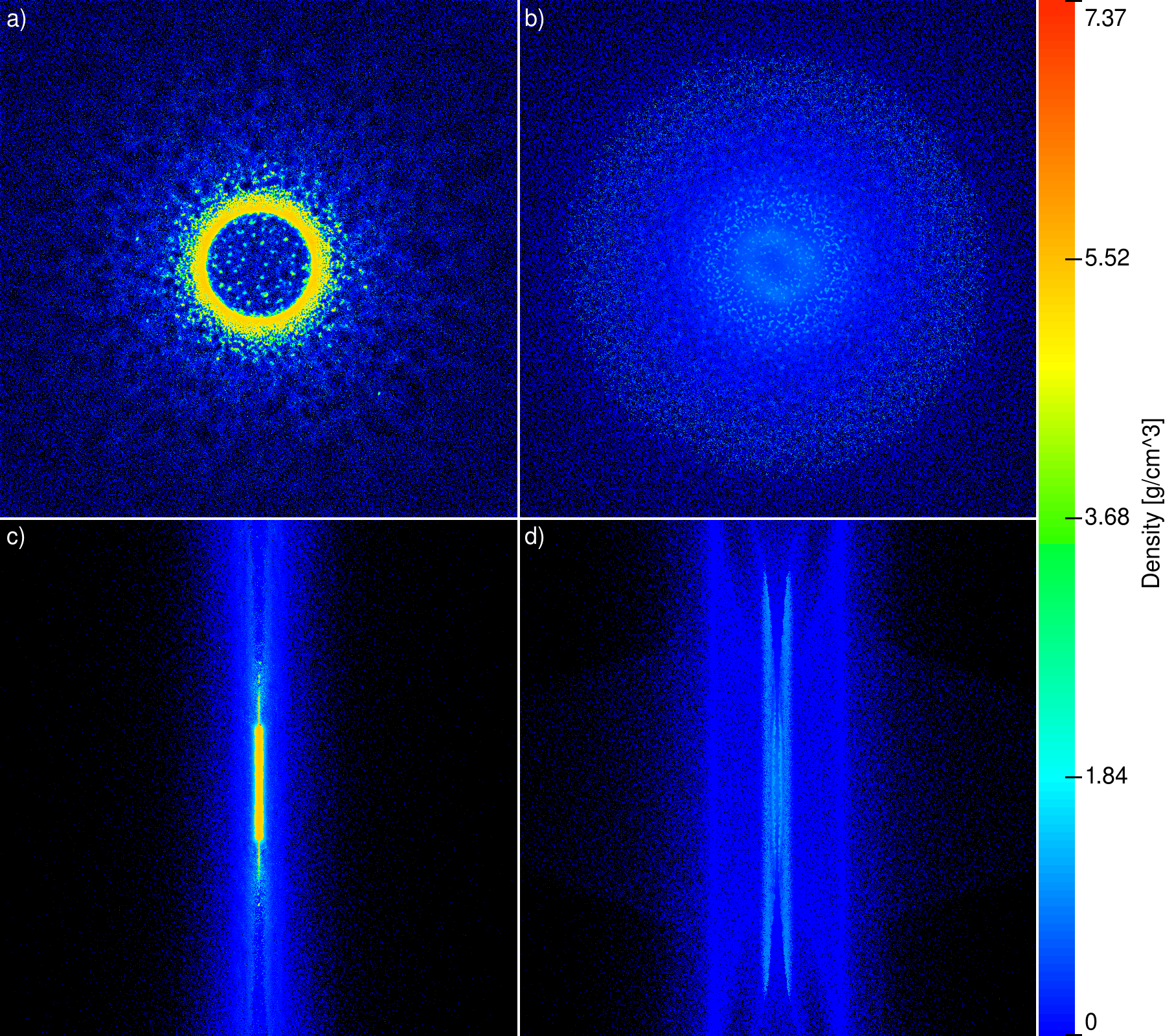}
    \caption{
    The ring-like structure that forms at higher resolutions in the head-on collisions. Shown is a snapshot of size \SI{10}{\Rearth} $\times$ \SI{10}{\Rearth} $\times$ \SI{10}{\Rearth} of a simulation using $1.3 \times 10^6$ particles \SI{2.3}{\hour} after the impact for ANEOS (left: a front view, c side view) and the Tillotson EOS (right: b front view, d side view). In both simulations the impact velocity is $2.67 v_{\textnormal{esc}}$. The colour corresponds to the density of the material ranging from \SI{0}{\gram\per\centi\meter\tothe3} (blue) to \SI{7.37}{\gram\per\centi\meter\tothe3} (red), the yellow material in the ring has a density of \SI{5.34}{\gram\per\centi\meter\tothe3}. All the material is in an expanded state and has a density below the reference density $\rho_0$. In case of the Tillotson EOS the remaining bound mass has extremely low densities and forms a central blob that consists of an iron/rock mixture. In the ANEOS simulation the outcome looks very different. The central part of the bound material forms a much denser, ring-like structure that is composed of iron originating from the proto-planetary cores. The iron particles are located at the solid/liquid-vapour phase boundary and have little pressure support. This in turn seems to affect convergence of \QRD.
    }
    \label{fig_Rings}
\end{figure*}

The head-on collisions result in a gravitationally bound remnant and a cloud of ejecta. The inferred value of \QRD for a given impact condition depends on the choice of EOS for all resolutions (see Figure \ref{fig_QDstarN-1_3}). For a given impact velocity the remaining bound remnant is up to \SI{23}{\percent} larger when ANEOS is used and the inferred value of \QRD differs by \SI{10.6}{\percent} to \SI{16.4}{\percent} between the two EOS. Therefore, if the colliding planets are modelled with ANEOS substantially higher impact energies are required for critical disruption. Since the colliding proto-planets have almost identical radii for both EOS (see Table \ref{tab_BodyProperties}), their gravitational binding energies differ by less than \SI{1.5}{\percent}. The observed difference in the stripped mass for given impact conditions is a result of how the two EOS model the interaction of the shock and rarefaction wave at the free surface (e.g., \citealt{stewartShockPhysicsGiant2020}). This in turn affects the particle velocity at the free surface and determines the amount of ejected material. The Tillotson EOS has a very simplistic treatment of the expanded, intermediate states and severely underestimates the sound speed compared to ANEOS. Due to the relatively low sound speed the particle velocity at the free surface and therefore the amount of ejected material is overestimated.

As in G15 the inferred value for \QRD also depends on the simulations resolution (see Figure \ref{fig_QDstarN-1_3}). For the Tillotson EOS we find that \QRD varies with the number of particles $N$ as
\begin{equation}
    Q_{\textnormal{RD}}^{*} \left( N \right) = a N^{-\frac{1}{3}} + b
\end{equation}
where $a$ and $b$ are fitting constants proposed in G15. We find $a=\SI{2.76e8}{\joule\per\kilo\gram}$ and $b=\SI{0.59e8}{\joule\per\kilo\gram}$. Therefore, as in G15 over $10^8$ particles are required to agree within \SI{1}{\percent} with the converged value $b$ in the limit of $N\rightarrow\infty$ particles. Even so, at such high resolutions simulations would probe fine details of the EOS, drawing into question any physical interpretation of the convergence using the Tillotson EOS. For lower resolutions, i.e., $N<10^5$, the corresponding simulations where ANEOS is used show the same behaviour and we find $a=\SI{3.31e8}{\joule\per\kilo\gram}$ and $b=\SI{0.65e8}{\joule\per\kilo\gram}$. However, for $N>10^5$ particles the inferred value of \QRD deviates from this trend, remains constant and seems to have converged to the asymptotic value of $\SI{7.2e7}{\joule\per\kilo\gram}$. While differences in \QRD due to the choice of EOS are expected, this sudden change in the convergence behaviour is surprising.

A more detailed investigation reveals an interesting difference between the two EOS during post-shock expansion of the material (see Figure \ref{fig_Rings}). For both EOS the gravitationally bound material at impact energies close to \QRD is entirely in the expanded states, i.e., has a density that is smaller than the reference density $\rho_0$ of the solid at zero compression. In the case of the Tillotson EOS, most of the material is in the intermediate expanded states region of the EOS. Independent of the simulations resolution it has an almost uniformly low density. The central, slightly denser region consists of an iron/granite mixture. In general, iron and granite are well mixed and the material shows little structure.

The simulations that were performed with ANEOS look very different. While most of the material has a similar low density as in the Tillotson EOS case, the central region is much denser. Below $10^5$ particles this material forms a central dense clump consisting of an iron-rock mixture. Once the resolution increases, an additional feature emerges in the formation of a dense ring, which later collapses under self gravity to form the core of the remnant. The inner, denser part (yellow in Figure \ref{fig_Rings}) is composed of iron originating from the colliding proto-planetary cores and is surrounded by a layer of rock (green in the same figure)\footnote{An animation showing the formation and evolution of this ring-like structure at a resolution of $1.3 \times 10^6$ particles can be found in the supplementary material of the journal.}. The iron particles that form the ring are located at the solid/liquid-vapour phase boundary where the pressure is constant with increasing density and therefore have little pressure support. We additionally investigate the collisions between very low mass, undifferentiated bodies at different resolutions following G15. Due to the lack of an iron core, no ring is formed in any simulation and we reproduce the scaling of \QRD found in G15 for both EOS. The gravitational potential generated by the ring seems to keep the surrounding material from expanding to lower densities contrary to what was observed in the case of the Tillotson EOS. This in turn promotes gravitational re-accumulation of material in a later phase following the collision and formation of this ring. Note that this ring is not observed in even the highest resolution Tillotson simulations. If the resolution is high enough to resolve the formation of the ring, the \QRD value has converged. If the EOS used to model iron does not model the solid/liquid-vapour phase transition or no iron core is present then convergence is as observed in the Tillotson case.

\section{Summary and conclusions}
\label{sec_conclusions}
In the present work we investigate how the choice of EOS and resolution conspire to affect the outcome of GI using state-of-the-art 3D hydro simulations. We compare the two most popular EOS, the Tillotson EOS and ANEOS, and determine two fundamental properties of the post-impact planet and their numerical convergence. During the last stage of planet formation collisions between similar mass proto-planets are common, therefore we consider two classes of impacts between differentiated, equal-mass bodies of \SI{0.5}{\Mearth}. First, we determine the rotation period of the merged planet in oblique collision involving initially rotating and non-rotating bodies for both EOS and different resolutions. Then we investigate in head-on collisions of non-rotating bodies how the critical specific impact energy \QRD required for catastrophic disruption depends on the choice of EOS and numerical resolution. In both cases we find that the outcomes are affected by the choice of EOS as well as the numerical resolution. This is consistent with prior work (e.g., \citealt{benzOriginMoonSingle1989,canupSimulationsLateLunarforming2004,gendaResolutionDependenceDisruptive2015,kegerreisPlanetaryGiantImpacts2019, reinhardtBifurcationHistoryUranus2020}).

Furthermore, we observe that the inferred values converge much faster with increasing resolution once the additional structures that emerge due to a more accurate and thermodynamically consistent treatment of the expanded states in ANEOS are resolved. In case of the Tillotson EOS such structures are not modelled by the EOS and therefore can also not be resolved even at ultra large resolutions. While it is not surprising, that the choice of the EOS or resolution of a simulation can affect the outcome of the collision, the effect on numerical convergence due to a physically more realistic EOS is a new finding that, to our knowledge, has not been reported in the literature before. In the ANEOS simulations the rotation period shows little resolution dependence and has essentially already converged when $10^4$ particles are used. Such a simulation requires little computational time even on a personal computer. Also \QRD converges at a resolution of $10^5$ particles for ANEOS while the Tillotson EOS simulations require over $10^8$ particles in order to agree within \SI{1}{\percent} with the converged value (in agreement with G15). However, the physical relevance of convergence is questionable in this context, as the Tillotson EOS used in the simulation no longer faithfully models the relevant processes at the resolved scales.

Since the observed differences in convergence of the rotation period depend on the amount of vapour generated in the collision, convergence could even further improve if an EOS like M-ANEOS \citep{meloshHydrocodeEquationState2007}, with a more consistent treatment of molecular vapour, is used. Another short-coming of ANEOS in it's standard form that is relevant in the context of this work is that it uses the same Debye thermal model for the solid and liquid phase which can result in wide deviations in entropy and temperature from experimental Hugoniot data and this in turn affects the amount of vapour generated in a collision \citep{stewartShockPhysicsGiant2020}. Additionally, inaccuracies in interpolation due to the interpolation method or the sampling of the EOS table with grid points could in principle affect convergence. Given that we find good agreement between direct EOS calls and interpolated data (see last paragraph of Section \ref{sec:interpolation}) we do not expect this to be the case in the present study but suggest to keep this possibility in mind when generating EOS tables. The parameter space in our simulations is limited and we therefore propose to investigate these findings in future simulations for different proto-planet masses and target to impactor mass ratios. Furthermore, investigating such impacts using different numerical schemes such as a finite volume code could provide further insight and validate our findings.

Our findings have profound implications for impact simulations. For example, accurately determining the planet's rotation period and angular momentum is crucial when investigating if GI on the ice giants can reproduce their tilt and rotation period (e.g., \citealt{kegerreisConsequencesGiantImpacts2018,kurosakiExchangeMassAngular2018,reinhardtBifurcationHistoryUranus2020,chauCouldUranusNeptune2021}). While the total angular momentum of the bound material, a key constraint in case of the Moon forming GI (e.g., \citealt{canupDynamicsLunarFormation2004}), shows little variation for the two compared EOS and different resolutions, the Earth's post-impact rotation period affects the tidal interaction with the Moon. This in turn affects the Moon's outward migration rate and the efficiency of AM removal due to evection resonances \citep{rufuTidalEvolutionEvection2020}, a key mechanism in reconciling high AM Moon forming impacts with observations (\citealt{canupFormingMoonEarthlike2012,cukMakingMoonFastSpinning2012,lockOriginMoonTerrestrial2018}). The impact conditions that lead to erosion of a proto-planet or smaller body are characterised by the value of \QRD and play a crucial role in determining many important properties of the final planetary system like the final number of planets, their mass and bulk composition. Applications in our Solar system include stripping Mercury's primordial mantle to explain it's large iron core (e.g., \citealt{benzOriginMercury2007,asphaugMercuryOtherIronrich2014,chauFormingMercuryGiant2018}) or the Asteroid Psyche's metal rich surface \citep{matterEvidenceMetalrichSurface2013}.

Future work should investigate, if a similar convergence behaviour can be observed for other attributes of the post-impact planet. One interesting quantity would be the disc mass, which was recently found to be non-convergent in ultra high resolution simulations in case of the Tillotson EOS \citep{hosonoUnconvergenceVerylargescaleGiant2017}. Another important aspect which has profound implications for the Moon forming impact is how the composition of the proto-satellite disc and mixing of target and impactor material depends on the choice of EOS and resolution.

Our findings highlight the importance of the interaction of the EOS and the numerical method in impact simulations and suggest that increased realism of the EOS in order to correctly model the resolved physics, e.g., an adequate treatment of phase mixtures and phase transitions in the expanded states, plays a key role when increasing the simulation's resolution.

\section*{Acknowledgements}
We thank Miles Timpe for providing the collision parameters for the oblique impacts and Alice Chau for helpful comments regarding the manuscript. We also thank the anonymous referee for valuable suggestions and comments that helped to improve the paper. The simulations were performed using the UZH HPC allocation on the Piz Daint supercomputer at the Swiss National Supercomputing Centre (CSCS). This work has been carried out within the framework of the National Centre of Competence in Research PlanetS, supported by the Swiss National Foundation.

\section*{Data Availability}
The data underlying this article are available in the Dryad Digital Repository, at \url{https://doi.org/10.5061/dryad.6q573n5zg}.

%%%%%%%%%%%%%%%%%%%% REFERENCES %%%%%%%%%%%%%%%%%%
\bibliographystyle{mnras}
\bibliography{main} % if your bibtex file is called example.bib

%%%%%%%%%%%%%%%%%%%%%%%%%%%%%%%%%%%%%%%%%%%%%%%%%%

%%%%%%%%%%%%%%%%% APPENDICES %%%%%%%%%%%%%%%%%%%%%
\appendix
\section{Calculation of the oblique collision trajectory}
\label{appendix_oblique_initial_condition}
The impact trajectory of the oblique collisions is defined by the asymptotic impact parameter $b_{\textnormal{inf}}$ and relative velocity $v_{\textnormal{inf}}$ at infinity. For all our oblique collision simulations, we set the asymptotic relative velocity as $v_{\textnormal{inf}}=0.5 v_{\textnormal{esc,sys}}$ where 
\begin{equation}
    v_{\textnormal{esc,sys}}=\sqrt{\frac{2G(M_{\textnormal{targ}}+M_{\textnormal{proj}})}{r_{\textnormal{crit}}}}
\end{equation}
is the escape velocity of the system and $r_{\textnormal{crit}}=r_{\textnormal{target}}+r_{\textnormal{proj}}$ the critical radius. The masses and radii of the two proto-planets are $M_{\textnormal{targ}}$, $M_{\textnormal{proj}}$, $r_{\textnormal{target}}$ and $r_{\textnormal{proj}}$ respectively.
For the asymptotic impact parameter we use $b_{\textnormal{inf}}=0.95 b_{\textnormal{max}}$ where
\begin{equation}
    b_{\textnormal{max}}=r_{\textnormal{crit}}\sqrt{1+\Theta}
\end{equation}
is the maximum asymptotic impact parameter that would lead to a collision where
\begin{equation}
    \Theta=\frac{v_{\textnormal{esc,targ}}^2}{v_{\textnormal{inf}}^2}
\end{equation}
is the Safronov number and
\begin{equation}
    v_{\textnormal{esc,targ}}=\sqrt{\frac{2GM_{\textnormal{targ}}}{r_{\textnormal{crit}}}}
\end{equation}
the escape velocity of the target. From these expressions the impact velocity
\begin{equation}
    v_{\textnormal{imp}}=\sqrt{v_{\textnormal{inf}}^2+v_{\textnormal{esc,targ}}^2}
\end{equation}
and impact parameter
\begin{equation}
    b_{\textnormal{imp}}=\frac{b_{\textnormal{inf}}v_{\textnormal{inf}}}{v_{\textnormal{imp}}r_{\textnormal{crit}}}
\end{equation}
are calculated. The initial positions and velocities of the two bodies are then given by
\begin{equation}
    r_{\textnormal{target}}= \begin{pmatrix}
	0\\0\\0
\end{pmatrix},\quad r_{\textnormal{projectile}}=\begin{pmatrix}
	b_{\textnormal{imp}}r_{\textnormal{crit}}\\-r_{\textnormal{crit}}\sqrt{1-b_{\textnormal{imp}}^2}\\0
\end{pmatrix}
\end{equation}
and
\begin{equation}
    v_{\textnormal{target}}=\begin{pmatrix}
	0\\0\\0
\end{pmatrix},\quad v_{\textnormal{projectile}}= \begin{pmatrix}
	0\\v_{\textnormal{imp}}\\0
\end{pmatrix}
\end{equation}
in the rest frame of the target body. In order to account for tidal deformation of the proto-planets prior to the collision their position and velocities are then evolved under mutual gravity until they have an initial separation of $r_{\textnormal{init}}=10r_{\textnormal{crit}}$. These modified positions and velocities are then transformed into the centre of mass frame and used as initial conditions for the collision simulation.

\section{Treatment of the negative pressure region in the expanded intermediate states of the Tillotson EOS}
\label{appendix_treatment_neg_pres_till}
In the expanded, cold states of the Tillotson EOS is a region where it returns a negative pressure attempting to model tensile forces in a solid \citep{meloshImpactCrateringGeologic1989}. Since this is clearly unphysical for a fluid, the pressure is commonly limited and either set to zero or a very small positive value in impact simulations. Negative pressures can also result in imaginary sound speed which is again unphysical and will cause undesired side effects in a hydro code. Therefore the sound speed in this region is also set to a positive minimum value.

In the expanded, intermediate states the Tillotson EOS interpolates in pressure between a cold, low-density solid/liquid and a vapour phase (e.g., \citealt{benzOriginMoonSingleimpact1986}). 
Depending on how the pressure limit is applied to this calculation the pressure and sound speed can differ for a given set of EOS parameters. In principle, one can either include the (potentially large) negative values in the calculation and limit the interpolated value or apply the limit to the pressure in the expanded, cold states before interpolation. Since the former results in extremely low pressures and sound speeds in large parts of the interpolation region \citep{stewartShockPhysicsGiant2020} and because we see no compelling reason to include unphysical, negative values in the calculation we do the latter in all our simulations. We set the pressure to zero if it would be negative and use the reference bulk sound speed as a minimum value for the sound speed \citep{reinhardtNumericalAspectsGiant2017,reinhardtBifurcationHistoryUranus2020}.

%%%%%%%%%%%%%%%%%%%%%%%%%%%%%%%%%%%%%%%%%%%%%%%%%%

% Don't change these lines
\bsp	% typesetting comment
\label{lastpage}
\end{document}